\documentstyle[12pt,epsf,titlepage]{article}
\begin{document}
\title{\bf {Fluctuation Effects and Multiscaling of the Reaction-Diffusion
Front for $A+B\rightarrow\emptyset$.}}
\author{Martin Howard and John Cardy$^{\dag}$}
\date{
{\small\noindent Department of Physics, Theoretical Physics, 1 Keble Road,
Oxford, OX1 3NP, U.K. \\}
{\small $^{\dag}$ and All Souls College, Oxford. \\}
{\bf PACS Numbers: 02.50.-r, 05.40.+j, 82.20.-w.\\}
%{\small Short Title: Fluctuations in the Reaction-Diffusion Front for
%$A+B\rightarrow\emptyset$.}
}\maketitle
\begin{abstract}
We consider the properties of the diffusion controlled reaction $A+B\rightarrow
\emptyset$ in the steady state, where fixed currents of $A$ and $B$ particles
are maintained at opposite edges of the system. Using
renormalisation group methods, we explicitly
calculate the asymptotic forms of the reaction front and particle
densities as expansions
in $(JD^{-1}|x|^{d+1})^{-1}$, where $J$ are the (equal) applied
currents, and $D$ the
(equal) diffusion constants. For the asymptotic densities
of the minority species, we find, in addition to the expected
exponential decay, fluctuation induced power law
tails, which, for $d<2$, have a universal form $A|x|^{-\omega}$,
where $\omega=5+O(\epsilon)$, and $\epsilon=2-d$.
A related expansion is derived for the reaction rate profile
$R$, where we find the asymptotic power law $R\sim B|x|^{-\omega -2}$.
For $d>2$, we
find similar power laws with $\omega=d+3$, but with non-universal
coefficients. Logarithmic corrections occur in $d=2$. These results
imply that, in the time dependent case,
with segregated initial conditions, the moments $\int |x|^{q}R(x,t)dx$
fail to satisfy simple scaling for $q>\omega+1$.
Finally, it is shown that the
fluctuation induced wandering of the position of the reaction front
centre may be neglected for large enough systems.
\end{abstract}

\section{Introduction}

Since the initial work of G\'alfi and R\'acz \cite{GaRa}, there has been
considerable interest in the kinetics of one and two species annihilation,
$A+A\rightarrow\emptyset$, and $A+B\rightarrow\emptyset$ [1--17, 23--27].
Most analytic and numerical studies have concentrated on the case of either
homogeneous initial conditions, or initially entirely segregated reactants.
Ben-Naim and Redner \cite{Red} were the first to study the case of a steady
state reaction interface, maintained by fixed particle currents
imposed at opposite edges of the system. Their
equations for the particle densities $a({\bf x},t)$ and $b({\bf x},t)$ were
\begin{eqnarray}
& & {\partial a\over\partial t}=D\nabla^2 a -\lambda ab \\
& & {\partial b\over\partial t}=D\nabla^2 b -\lambda ab,
\end{eqnarray}
with diffusion constant $D$, reaction rate constant $\lambda$, and with
the boundary conditions:
\begin{equation}
J=-D\partial_x a|_{x=-L} \quad 0=-D\partial_x b|_{x=-L} \quad 0=-D
\partial_x a|_{x=L} \quad -J=-D\partial_x b|_{x=L}.
\end{equation}
These equations are asymptotically soluble analytically, giving
\begin{equation}
{a\brace b}\sim (J/D)|x|{\theta(-x)\brace\theta(x)}+(const.)\left(J^2\over
\lambda D\right)^{1/3}\left({\lambda J\over D^2}\right)^{-1/12}|x|^{-1/4}
e^{-{2\over 3}(\lambda J/D^2)^{1/2}|x|^{3/2}},
\end{equation}
where $\theta(x)$ is the Heaviside step function. The relations $w\sim
J^{-1/3}$, where $w$ is the reaction front width, and $c\sim J^{2/3}$,
where $c$ is the particle concentration in the reaction
zone, are also derived in \cite{Red}. However,
implicit in their formulation is the `mean field' like assumption,
$\langle ab\rangle\propto\langle a\rangle\langle b\rangle$, which will
no longer be adequate below the critical dimension, due to fluctuations.
Cornell and Droz \cite{CD} have given an argument for the upper critical
dimension of the system (leading to $d_c=2$), as well as
performing numerical simulations. On the basis of these, and mean field
analysis, they have proposed scaling forms for $a$,$b$ and the reaction front
$R$, which are postulated to be valid both above and below the
critical dimension in the scaling limit $w\rightarrow\infty$ (or
$J\rightarrow 0$):
\begin{equation}
R={J\over w}S\left({x\over w}\right) \qquad a={wJ\over
D}A\left({x\over w}\right) \qquad b={wJ\over D}B\left({x\over
w}\right). \label{scale}
\end{equation}
In other words, the profiles are characterised by a single length scale
$w$, which itself is suggested in \cite{CD} to vary as $w\sim J^{-1/2}$ in
$d=1$, and $w\sim J^{-1/3}$ for $d\geq 2$ in the scaling limit. Cardy
and Lee \cite{Lee3} have given RG arguments which support this conclusion.
However, we defer further discussion, especially with regard to the
presence of multiscaling, until section 6.

In this paper, we present the results of the first renormalisation group
calculation for the asymptotic properties of the densities and
reaction front in the steady state, which systematically takes into
account the effect of fluctuations in the stochastic particle
dynamics. Previously, the RG had
been used to study the late time behaviour of reactions with
homogeneous initial conditions (see \cite{Lee} and references therein).
Our calculational framework will bear considerable similarities with
\cite{Lee}. The basic plan is to map the microscopic dynamics, in the
form of a master equation, onto a quantum field theory. This theory is
then renormalised (for $d\leq 2$) by the introduction of a renormalised
coupling, which is shown to have a stable fixed point of order $\epsilon$.
We then group the Feynman diagrams into sets whose sums give a
particular order of the renormalised coupling constant. It will be demonstrated
that this grouping is given by the number of loops. These diagrams
may then be evaluated (asymptotically) and the Callan-Symanzik solution
used, to obtain perturbative expansions for the densities and reaction front.
Note that for $d>2$ no renormalisation is necessary and the diagrams may be
evaluated directly.

We now present our results for the asymptotic forms of the densities
and reaction front profile. It will be shown that at zero loops we find
a stretched exponential dependence which
we include in the following summary, even though we expect its effects
to be overwhelmed by leading and subleading power law terms.
In addition, for $d<2$, we do not rule out the possibility of
logarithms in higher order terms summing to give a modification to the
leading power law given below (which results from the
straightforward evaluation of the one loop contributions).
So we find asymptotically as $|x|\rightarrow\infty$ for $d<2$:
\begin{equation}
{\langle a\rangle\brace\langle b\rangle}=(J/D)|x|{\theta(-x)\brace
\theta(x)}+A_1|x|^{(7-5d)/12}e^{-A_2|x|^{(d+1)/2}}+A_3|x|^{-5+2\epsilon}+\ldots
\end{equation}
\begin{equation}
R=\lambda\langle ab\rangle=A_4|x|^{(7d-5)/12}e^{-A_2|x|^{(d+1)/2}}
+A_5|x|^{-7+2\epsilon}+\ldots
\end{equation}
where
\begin{equation}
A_1=0.3787{(J/D)^{7/12}\over (4\pi\epsilon)^{5/12}}\qquad A_2={2\over
3}(4\pi\epsilon)^{1/2}(J/D)^{1/2}\qquad A_3={1\over
32\pi^2(J/D)\epsilon}
\end{equation}
\begin{equation}
A_4=0.3787(J/D)^{19/12}{(4\pi\epsilon)^{7/12}\over (9/D)}(d+1)^2\qquad
A_5={(2d+1)(2d+2)\over 32\pi^2(J/D^2)\epsilon};
\end{equation}
for $d=2$:
\begin{equation}
{\langle a\rangle\brace\langle b\rangle}=(J/D)|x|{\theta(-x)\brace\theta(x)}
+B_1(\ln|x|)^{5\over 12}|x|^{-{1\over 4}}e^{-B_2(\ln |x|)^{-1/2}|x|^{3/2}}+B_3
|x|^{-5}\ln|x| + \ldots
\end{equation}
\begin{equation}
R=B_4(\ln |x|)^{-7/12}|x|^{3/4}e^{-B_2(\ln |x|)^{-1/2}|x|^{3/2}}
+B_5|x|^{-7}\ln |x|+\ldots
\end{equation}
where
\begin{equation}
B_1=0.3787{(J/D)^{7/12}\over (4\pi)^{5/12}}\qquad B_2={2\over
3}(4\pi(J/D))^{1/2}\qquad B_3={(J/D)^{-1}\over 32\pi^2}
\end{equation}
\begin{equation}
B_4=0.3787(4\pi)^{7/12}D(J/D)^{19/12}\qquad B_5={15(J/D)^{-1}\over 16\pi^2/D};
\end{equation}
and finally, for $d>2$:
\begin{equation}
{\langle a\rangle\brace\langle b
\rangle}=(J/D)|x|{\theta(-x)\brace\theta(x)}
+C_1|x|^{-1/4}e^{-C_2|x|^{3/2}}+C_3|x|^{-d-3}+\ldots
\end{equation}
\begin{equation}
R=C_4|x|^{3/4}e^{-C_2|x|^{3/2}}+C_5|x|^{-d-5}+\ldots
\end{equation}
where
\begin{equation}
C_1=0.3787{(J/D)^{7/12}\over (\lambda/D)^{5/12}}\qquad C_2={2\over
3}(\lambda J/D^2)^{1/2}
\end{equation}
\begin{equation}
C_3=(\lambda J/D^2)^{-1}2^{-1-d}\pi^{-(d+1)/2}(d-1)\Gamma\left({d-1\over
2}\right) \qquad C_4=0.3787\lambda{(J/D)^{19/12}\over (\lambda/D)^{5/12}}
\end{equation}
\begin{equation}
C_5=\lambda(\lambda/D)^{-2}(J/D)^{-1}2^{-1-d}\pi^{-(d+1)/2}\Gamma\left({
d-1\over 2}\right)(d-1)(d+3)(d+4).
\end{equation}
The layout of this paper is as follows. In section 2, the system is
defined using a master equation, which is then mapped to a second quantised
representation, and then to a field theory. In section 3, we present two
related field theories and derive the form of their Green functions. The
renormalisation of the theory is also addressed.
The calculations for the densities and reaction front are presented
in section 4, for $d<d_c$, $d=d_c$, and $d>d_c$. The separate problem of the
fluctuations in the position of the centre of the reaction front is
presented in section 5. A discussion of these results and comparisons
with the available data from simulations are given in section 6, where
we also argue for the presence of multiscaling in the system.

\section{The Model}

We consider a model where A and B particles are moving diffusively on a
hypercubic lattice, with lattice constant $l$. There is some
probability of mutual annihilation whenever an A and a B particle meet on a
lattice site. In addition, particles of type A are added at a constant rate to
lattice sites on the hypersurface $x=-L$, and particles of type B are
similarly added to sites at $x=L$. In other words, opposing currents of A and
B particles are maintained at opposite edges of the system. The two
hypersurfaces $x=\pm L$ mark the boundaries of the system beyond which the
particles are not permitted to move. The model is defined by
a master equation for $P(\{n,m\},t)$, the
probability of particle configuration $\{n,m\}$ occurring at time $t$. Here
$\{n,m\} = (n_1,n_2, \ldots ,n_N,m_1,m_2, \ldots ,m_N)$, where $n_i$ is the
occupation number of the A particles, and $m_i$ the occupation number of the B
particles, at the $i$th lattice site. The appropriate master equation is
\begin{eqnarray}
& & \qquad{\partial \over \partial t}P(\{n,m\},t) =
\nonumber\\& & {D \over l^{2}} \sum_{i,e} \{
(n_e+1)P(\dots n_i-1,n_e+1, \dots ,\{m\},t)-n_iP(\{n,m\},t)\} \qquad \qquad
\nonumber \\
& &+ {D \over l^{2}} \sum_{i,e}\{(m_e+1)P( \{n\}, \dots m_i-1,m_e+1, \dots,t)
-m_iP(\{n,m\},t) \} \nonumber \\ & &+ \lambda \sum_i [(n_i+1)(m_i+1)P( \dots
n_i+1,m_i+1 \dots ,t) - n_i
m_iP( \dots n_i,m_i \dots ,t)] \nonumber \\ & &+ R \{ P( \dots,
n_{-L}-1,\{m\},t)  -  P(\dots, n_{-L}, \{m\},t)\} \nonumber \\  & &+
R \{ P( \{n\}, m_L-1 \dots ,t) -  P( \{n\}, m_L \dots ,t)\},
\end{eqnarray}
where $i$ is summed over lattice sites, and $e$ is summed over the nearest
neighbours of $i$. The first, second, and third lines of the equation describe
diffusion
of the A and B particles respectively (with equal diffusion constants $D$),
whilst the fourth line describes their annihilation within the system (with
rate constant $\lambda$). The final four terms are due to the addition of
A and B particles at the edges of the system at a rate $R$, corresponding to
the maintenance of steady particle currents.

The master equation can be mapped to a second quantised form, following a
standard procedure developed by Doi \cite{Doi} and Peliti \cite{Peliti}, and
as described by
Lee \cite{Lee}. In brief, in terms of the creation/annihilation operators
$a,a^{\dag},b,b^{\dag}$
which are introduced at each lattice site, the time evolution operator for
the system is
\begin{equation}
\hat H = -{D \over l^2} \sum_{i,e} \{a_i^{\dag}(a_e-a_i) + b_i^{\dag}(b_e-b_i)
\} - \lambda \sum_i(1 - a_i^{\dag} b_i^{\dag})a_ib_i - R\{a_{-L}^{\dag}
-1 + b_{L}^{\dag}-1\}.
\end{equation}
This can now be mapped onto a path integral, in which $a,\hat a,b,\hat b$ are
replaced by continuous c-number fields, with action (up to a constant)
\begin{eqnarray}
& & S = \sum_i \left( \int_{-\infty}^t dt \left\{ \hat a_i \dot a_i +
\hat b_i \dot b_i - {D \over l^2} \left[\hat a_i \sum_e(a_e-a_i) +\hat b_i
\sum_e (b_e-b_i)\right] \right.\right. \nonumber \\ & & \left.\left.\qquad
\qquad-\lambda ( 1-\hat a_i\hat b_i)a_i b_i \right\} - a_i(t) - b_i(t) \right),
\end{eqnarray}
where $a_i(t)$ and $b_i(t)$ are due to the projection state (see \cite{Lee}),
provided that
\begin{equation}
R = {J \over l}={D \over l^2}(a_o - a_{-L}) \qquad 0={D \over l^2}
(b_o-b_{-L})
\end{equation}
at $-L$, and
\begin{equation}
R = {J \over l}={D \over l^2}(b_o-b_L) \qquad 0={D \over l^2}
(a_o-a_L)
\end{equation}
at $+L$. Here the sites $-L$, $L$ are at the edges of the system, with
site $o$ being immediately outside $L$ or $-L$.
Taking the continuum limit of this action, we arrive at
\begin{eqnarray}
& & S[\hat a,a,\hat b,b,t] = \int \left( dx d^{d-1}y \int_{-\infty}^t
dt\left\{ \hat a (
\partial_t - D \nabla^2) a + \hat b ( \partial_t - D \nabla^2)b
\right.\right. \nonumber\\ & & \left.\left.\qquad\qquad\qquad\qquad\qquad\qquad
\qquad- \lambda_0 (1-\hat a \hat b)ab \right\} - a(t) -
b(t) \right).
\end{eqnarray}
subject to the conditions
\begin{equation}
-J = -D \partial_x b \qquad 0= -D \partial_x a
\end{equation}
at $+L$, and
\begin{equation}
J=-D \partial_x a \qquad 0=-D \partial_x b
\end{equation}
at $-L$. Here $y$ are the coordinates for directions perpendicular to
the applied currents. These conditions may be made explicit in
the action by including a
pair of delta functions:
\begin{eqnarray}
& & S=\int \left( dx d^{d-1}y\int_{-\infty}^t dt \left\{ \hat a(\partial_t
-D\nabla^2)a +\hat b(\partial_t -D\nabla^2)b \right.\right. \\ & &
\left.\left.\qquad\quad - \lambda_0(1-\hat a \hat b)ab -\hat a J\delta(x+L)
- \hat b J \delta(x-L)\right\} -a(t)-b(t)\right). \nonumber
\end{eqnarray}
If we make the substitutions $ \hat a = 1+ \bar a$ and $ \hat b=1+\bar b$,
then the action becomes (up to a constant)
\begin{eqnarray}
& & S=\int dx d^{d-1}y dt \left[ \bar a(\partial_t-D\nabla^2)a+
\bar b (\partial_t
-D\nabla^2)b +\lambda_0 \bar a ab + \lambda_0 \bar b ab \right. \nonumber\\
& & \left. \qquad \qquad\qquad\qquad+\lambda_0
\bar a \bar b ab - \bar a J \delta(x+L)-\bar b J\delta(x-L) \right].
\end{eqnarray}
If we integrate over the $\bar a$ and $\bar b$ fields, and neglect the
$\bar a \bar b ab$ term, we obtain the classical (mean field) equations
\begin{eqnarray}
( \partial_t -D\nabla^2)a +\lambda_0 ab-J\delta(x+L)=0 \\
( \partial_t -D\nabla^2)b +\lambda_0 ab-J\delta(x-L)=0.
\end{eqnarray}
On the further conditions that no particle annihilation occurs at the
edges of the system, and that $\nabla a=0$ and $\nabla b=0$ outside,
integrating the first equation from $-L-\epsilon$ to $-L+\epsilon$ and
the second from $L-\epsilon$ to $L+\epsilon$ in the limit $\epsilon
\rightarrow 0$ gives the required boundary conditions.

As the diffusion constant exhibits no singular behaviour in the renormalisation
of the theory, it is convenient to absorb it into a rescaling of time, as in
\cite{Lee}. Defining $\bar t=Dt$, $\bar \lambda=\lambda_0 D^{-1}$, and
$ \bar J=J D^{-1}$, and introducing the fields $\phi={1 \over 2}(a+b)$ and
$\psi={1 \over 2} (a-b)$, we have
\begin{eqnarray}
& & S=\int dx d^{d-1}y d\bar t\left[2\bar\phi(\partial_{\bar t}-\nabla^2)\phi
+2\bar\psi(
\partial_{\bar t}-\nabla^2)\psi
+2\bar\lambda\bar\phi(\phi^2-\psi^2)+ \right. \label{a1}\\ & & \left. \bar
\lambda(\bar\phi^2-\bar\psi^2)(\phi^2-\psi^2)
-\bar J\bar\phi[\delta(x+L)+\delta(x-L)]-
\bar J\bar\psi[\delta(x+L)-\delta(x-L)]\right]. \nonumber
\end{eqnarray}
Consequently, the new classical equations of the steady state are
\begin{eqnarray}
\partial_x^2 \psi_c +{1\over 2}\bar J[\delta(x+L)-\delta(x-L)]=0 \\
\partial_x^2 \phi_c - \bar\lambda(\phi_c^2-\psi_c^2) +{1\over 2}\bar
J[\delta(x+L)+\delta(x-L)]=0. \label{phc}
\end{eqnarray}
The appropriate solution for $\psi_c$ is just $-(\bar J/2)x$ (for
$|x|\leq L$),
whilst substituting $\phi_c= (\bar J/2)|x| + u$ into the second equation gives
asymptotically the Airy equation for $u$, as noted in \cite{Red}.
Asymptotically one finds
\begin{equation}
\phi_c={J\over 2D}|x| + 0.3787\left({J^2\over D\lambda}\right)^{1/3}
\left({\lambda
J\over D^2}\right)^{-1/12}|x|^{-1/4} e^{-(2/3)(\lambda J
D^{-2})^{1/2}|x|^{3/2}}+\ldots, \label{phic}
\end{equation}
where the constant was determined numerically.

So far the quartic term in the action has been neglected, with the result that
the simple mean field results have been recovered. However, we can take into
account the non-classical term by including Gaussian noise in the equations
for $\phi$ and $\psi$, leading to equations which are exact.
This modification can be derived by replacing the
quartic piece in the action by a noise variable, integrating over the
noise distribution, and demonstrating that this recovers the original term.
Observing that
\begin{equation}
\int_{-\infty}^{\infty} d\eta_{\phi} e^{\bar \phi \eta_{\phi}} e^{-\eta_
{\phi}^2/[4\bar\lambda(\phi^2-\psi^2)]} \sim e^{\bar\lambda \bar \phi^2(
\phi^2-\psi^2)},
\end{equation}
and
\begin{equation}
\int_{-\infty}^{\infty} d\eta_{\psi} e^{\bar \psi \eta_{\psi}} e^{-\eta_
{\psi}^2/[4\bar\lambda(\phi^2-\psi^2)]} \sim e^{-\bar\lambda \bar \psi^2(
\phi^2-\psi^2)},
\end{equation}
where $\eta_{\phi}$ and $\eta_{\psi}$ are complex Gaussian noise
variables with an appropriate phase, we see that the steady state
equations may be written as
\begin{eqnarray}
& & \partial_x^2 \psi +{1\over 2}\bar J[\delta(x+L)-\delta(x-L)]+\eta_\psi=0
\label{ee1}\\ & & \partial_x^2 \phi - \bar\lambda(\phi^2-\psi^2) + {1\over 2}
\bar
J[\delta(x+L)+\delta(x-L)] + \eta_\phi=0. \label{ee2}
\end{eqnarray}
Clearly we have lost the simple interpretation of the $a$ and
$b$ fields as being the local densities of A and B particles, as now each of
the above equations includes a (generally) complex noise term. Nevertheless,
we can still interpret $\langle\psi\rangle$ and $\langle\phi\rangle$ as being
averaged densities, which also satisfy
\begin{equation}
\partial_x^2\langle\psi\rangle+{1\over 2}\bar
J[\delta(x+L)-\delta(x-L)]=0
\end{equation}
\begin{equation}
\partial_x^2 \langle\phi\rangle - \bar\lambda\langle \phi^2-\psi^2 \rangle
+{1\over 2}\bar J[\delta(x+L)+\delta(x-L)] =0. \label{ptr}
\end{equation}
The second equation will be used later on to relate a perturbation expansion
for $\langle \phi\rangle$ to one for $\langle \phi^2-\psi^2\rangle$.

Finally, we give the natural canonical dimensions for the various quantities
appearing in the action, noting that the coupling becomes dimensionless at
the postulated value of the critical dimension \cite{CD}:
\begin{equation}
[\bar t]=k^{-2} \qquad [a,b]=k^{d} \qquad [\bar a,\bar b]=k^0 \qquad [\bar
\lambda]=k^{2-d} \qquad [\bar J]=k^{d+1}.
\end{equation}

\section{Field Theory Formulations}

In what follows it will be convenient to develop two parallel field theories -
one given by the action already described in (\ref{a1}), and another to be
described below, formed by writing $\phi = \phi_c + \phi_1$ and $\psi = \psi_c
+ \psi_1$. In particular, whilst the second theory is more useful for
calculations, the cancellation of divergences after
renormalisation, and the identification of leading terms in an expansion in
powers of the coupling constant, are easier to see in the first
theory.

\subsection{Propagators and Vertices}

The propagators for the first of the two theories described above
(which we shall call Field Theory I) are (from (\ref{a1}))
\begin{equation}
G_{\phi\bar\phi}(k,\bar t)=G_{\psi\bar\psi}(k,\bar t)={1\over 2}e^{-k^2\bar t},
\end{equation}
in $(k,\bar t)$ space. In $(k,s)$ space, where a Laplace transform of time has
been performed, we have
\begin{equation}
G_{\phi\bar\phi}(k,s)=G_{\psi\bar\psi}(k,s)={1/2 \over k^2+s}.
\end{equation}
The vertices are shown in figure 1, where the $\phi$ propagators are
represented by solid lines, and $\psi$ propagators by dotted lines.

For Field Theory II, we split the $\phi$ and $\psi$ fields into
their classical and non-classical components, which leads to a modified action
\begin{eqnarray}
& & S =\int dx d^{d-1}y d\bar t \left\{ 2\bar\phi(\partial_{\bar t}-\nabla^2
+2\bar\lambda\phi_c)\phi_1+ 2\bar\psi
(\partial_{\bar t} -\nabla^2)\psi_1 \right.\\& &\left.
+2\bar\lambda\bar\phi(\phi_1^2-\psi_1^2
-2\psi_c\psi_1)+\bar\lambda(\bar\phi^2-\bar\psi^2)
(\phi_c^2+2\phi_c\phi_1+
\phi_1^2-\psi_c^2-2\psi_c\psi_1-\psi_1^2)\right\}, \nonumber
\end{eqnarray}
where the classical equations have been used to simplify its form somewhat.
We can now substitute for the exact value of $\psi_c = -(\bar J/2)x$ and for
the functional form of the $\phi_c$ field (from (\ref{phc}))
\begin{equation}
\phi_c = (\bar J^2/\bar\lambda)^{1/3} f[(\bar\lambda\bar J)
^{1/3}x].
\end{equation}
If we also make the rescaling in the action of
\begin{equation}
\tilde x = (\bar\lambda\bar J)^{1/3} x \quad \tilde y=(\bar\lambda\bar
J)^{1/3}y \quad \tilde t=(\bar\lambda\bar J)^{2/3}\bar t,
\end{equation}
then it is transformed to
\begin{eqnarray}
& & S=\int d\tilde x d^{d-1}\tilde y d\tilde t (\bar\lambda\bar
J)^{-{d\over 3}}\left[2\bar
\phi(\partial_{\tilde t}-\tilde\nabla^2+2f(\tilde x))\phi_1+2\bar\psi
(\partial_{\tilde t}-\tilde\nabla^2)\psi_1 \right.\nonumber\\
& &\left.\qquad\qquad\qquad+2(\bar\lambda/\bar J^2)^{1/3}\bar\phi
(\phi_1^2-\psi_1^2)
+2\bar\phi\tilde x \psi_{1} \right.\nonumber\\
& &\left.+(\bar\phi^2-\bar\psi^2)\left(
(\bar J^2/\bar\lambda)^{1\over 3}h(\tilde x) +2f(\tilde x)\phi_1
+\tilde x \psi_1 +(\bar\lambda/\bar J^2)^{1\over 3}(
\phi_1^2-\psi_1^2)\right)\right],
\end{eqnarray}
where
\begin{equation}
h(\tilde x)= [f(\tilde x)]^2 - {1\over 4}\tilde x^2,
\end{equation}
which is essentially just the classical profile of the reaction front. The
form of the propagators is now
\begin{equation}
G_{\phi_1\bar\phi}={1\over 2}(\bar\lambda\bar J)^{d/3}G(\tilde x,
\tilde x',{\bf \tilde y,\tilde y'},\tilde t),
\end{equation}
where
\begin{equation}
[\partial_{\tilde t}-\tilde\nabla^2+2f(\tilde x)]G(\tilde x,\tilde x',
{\bf \tilde y,\tilde y'},\tilde t)=
\delta(\tilde x-\tilde x')\delta({\bf \tilde y-\tilde y'})\delta
(\tilde t-\tilde t'); \label{pgf}
\end{equation}
and
\begin{equation}
G_{\psi_1\bar\psi}(\tilde k,\tilde s)={1\over 2}(\bar\lambda\bar J)
^{d/3}{1\over\tilde k^2 +\tilde s},
\end{equation}
in $(\tilde k,\tilde s)$ space, or
\begin{equation}
G_{\psi_1\bar\psi}(\tilde x-\tilde x',\tilde k_{\perp},\tilde s)
=(\bar\lambda\bar J)^{d/3}{e^{-(\tilde k_{
\perp}^2+\tilde s)^{1/2}|\tilde x-\tilde x'|}\over 4(\tilde k_{\perp}
^2+\tilde s)^{1/2}},
\end{equation}
in $(\tilde x,\tilde k_{\perp},\tilde s)$ space, where the
perpendicular directions are defined to
be those perpendicular to the applied currents.
Unfortunately the equation for $G$ (\ref{pgf}) is too hard to solve exactly,
as we do not have an analytic form for $\phi_c$. Consequently we must rely
on the approximation $f(\tilde x) \sim{1\over 2}|\tilde x|$, valid at
large $|\tilde x|$,
in order to
make the equation tractable. If we also Laplace transform time, and Fourier
transform to momentum space for spatial dimensions perpendicular to the
applied currents, then we obtain
\begin{equation}
\left(\tilde s+ \tilde k_{\perp}^2-\partial_{\tilde x}^2+|\tilde x|\right)
G(\tilde x,\tilde x',\tilde k_{\perp},\tilde s)=\delta(\tilde x-\tilde x').
\end{equation}
This has the solution (for $\tilde x'>0$, and accurate for large
$|\tilde x|$ i.e. when  $|x|\gg (\bar\lambda\bar J)^{-1/3}$):
\begin{equation}
G(\tilde x,\tilde x',\tilde k_{\perp},\tilde s)=
\cases{\alpha Ai[-\tilde x +\tilde k_{\perp}^2 +\tilde s]& when
$\tilde x\leq 0$ \cr
\beta Ai[\tilde x+\tilde k_{\perp}^2+\tilde s]+\gamma Bi[\tilde x+\tilde
k_{\perp}^2+\tilde s]& when $0\leq \tilde x\leq \tilde x'$ \cr
\delta Ai[\tilde x+\tilde k_{\perp}^2 +\tilde s]& when $\tilde x\geq
\tilde x'$.\cr}
\end{equation}
Considering the boundary conditions at $\tilde x'$ (continuity in $G$ and a
discontinuity in its derivative), we have
\begin{eqnarray}
& & \beta Ai(\tilde x'+\tilde k_{\perp}^2+\tilde s)+\gamma Bi(\tilde x'+
\tilde k_{\perp}^2+\tilde s)=\delta Ai(\tilde x'+\tilde k_{\perp}^2+\tilde s)
\label{Ai} \\
& & \beta Ai'(\tilde x'+\tilde k_{\perp}^2+\tilde s)+\gamma Bi'(\tilde x'
+\tilde k_{\perp}^2+\tilde s)-\delta Ai'(\tilde x'+\tilde k_{\perp}^2
+\tilde s) =1.
\end{eqnarray}
These equations can be solved for $\gamma$ with the result that $\gamma =
\pi Ai(\tilde x'+\tilde k_{\perp}^2+\tilde s)$. The final boundary
condition ($G\rightarrow 0$ as $\tilde x\rightarrow -\infty$) will (in
principle) give a further relation between $\beta$ and $\gamma$, as
well as specifying $\alpha$. But to use this condition we need to know
the behaviour of $G$ in regions near $0$, where our asymptotic
approximation breaks down. Consequently we must rely on numerical
solutions, which reveal that for our purposes we may neglect the $\beta
Ai$ term in (\ref{Ai}). Solving for $\delta$, we obtain:
\begin{equation}
G(\tilde x,\tilde x',\tilde k_{\perp}^2,\tilde s)=
\cases{ \pi Ai(\tilde x'+\tilde k_{\perp}^2+\tilde s) Bi(\tilde x+\tilde
k_{\perp}^2+\tilde s) & for $0\ll \tilde x\leq\tilde x'$ \cr
\pi Bi(\tilde x'+\tilde k_{\perp}^2+\tilde s) Ai(\tilde x+\tilde k_{\perp}
^2+\tilde s) & for $\tilde x\geq\tilde x'\gg 0$.}
\end{equation}
We can now use the asymptotic form of the Airy function \cite{Han} to
simplify these expressions further:
\begin{equation}
Ai(z)\sim {1\over 2\sqrt{\pi}}z^{-1/4}e^{-(2/3)z^{3/2}}, \qquad Bi(z) \sim
{1\over \sqrt{\pi}}z^{-1/4}e^{(2/3)z^{3/2}}.
\end{equation}
Hence, for $0\ll \tilde x<\tilde x'$,
\begin{equation}
G = {1\over 2}(\tilde x'+\tilde k_{\perp}^2+\tilde s)^{-1/4}(\tilde x
+\tilde k_{\perp}^2
+\tilde s)^{-1/4} e^{-(2/3)[(\tilde x'+\tilde k_{\perp}^2+\tilde s)^{3/2}-(
\tilde x+\tilde k_{\perp}^2+\tilde s)^{3/2}]},
\end{equation}
with a similar expression for $\tilde x>\tilde x'\gg 0$. For $\tilde x
\approx \tilde x'\gg 0$, we may
expand the terms inside the exponential, to obtain
\begin{equation}
G_{\phi_1\bar\phi}(\tilde x,\tilde x',\tilde k_{\perp},\tilde s) = {1\over
4}(\bar\lambda\bar J)^{d/3}(\tilde x+
\tilde k_{\perp}^2+\tilde s)^{-1/2} e^{-(\tilde x+
\tilde k_{\perp}^2 +\tilde s)^{1/2}|\tilde x-\tilde x'|}.
\end{equation}
The important point to notice here is that for $\tilde x$ sufficiently
close to $\tilde x'$, the Green function decays only as a power law.

Finally, we note that each occurrence of a propagator is associated with a
factor of $(\bar\lambda\bar J)^{d/3}$. If we also
extract a factor of $(\bar\lambda\bar J)^{-(d+2)/3}$ from each vertex,
then we can use the vertices shown in figure 2, provided we multiply
any given diagram by a factor of
\begin{equation}
(\bar\lambda\bar J)^{{1\over 3}pd-{v\over 3}(d+2)},
\end{equation}
where $p$ is the number of propagators and $v$ is the number of vertices.
Again, in figure 2, $\phi$ propagators are solid lines and $\psi$
propagators are dotted lines. Note the simple form of the vertices (h)
to (m).

\subsection{Renormalisation}

The renormalisation of the theory proceeds in a similar vein to that
described in \cite{Lee} -  our field theory differs only in the nature of the
boundary conditions. Again the only renormalisation required is coupling
constant renormalisation, as the set of vertices for Field Theory I allows
no diagrams which dress the propagator. Hence we have no field
renormalisation and the bare propagators are the full propagators for the
theory.

\subsubsection{Renormalisation of the Coupling}

The temporally extended vertex function for $A + B
\rightarrow \emptyset$ is given by the sum of diagrams
shown in figure 3. This sum may be calculated exactly, as done in
\cite{Lee2} (remembering extra factors of two resulting from the
presence of two different types of propagator):
\begin{equation}
\bar \lambda(k,s)={\bar\lambda \over 1+{1\over 2}\bar\lambda B_2 \Gamma(
\epsilon/2)(s+{1\over 2}k^2)^{-\epsilon/2}},
\end{equation}
where $B_2 = 2/(8\pi)^{d/2}$, and $\epsilon=2-d$. However, as we are now in
the time independent state, we take $s=0$, leading to
\begin{equation}
\bar\lambda(k)={\bar\lambda\over 1+{1\over 2}\bar\lambda B_2\Gamma(\epsilon
/2)2^{\epsilon/2}k^{-\epsilon}}.
\end{equation}
The vertex function can now be used to define the renormalised coupling, with
$k=\kappa$ as the normalisation point (differing from \cite{Lee}). So we have
\begin{equation}
g_R=\kappa^{-\epsilon} \bar\lambda(k)|_{k=\kappa} \qquad
g_0=\kappa^{-\epsilon} \bar\lambda,
\end{equation}
for the dimensionless renormalised and bare couplings respectively.
The $\beta$ function
is defined by
\begin{equation}
\beta(g_R)\equiv\kappa{\partial\over \partial\kappa}g_R
=-\epsilon g_R +{1\over 2}
\epsilon g_R^2 B_2\Gamma(\epsilon/2)2^{\epsilon/2}, \label{bf}
\end{equation}
and we have a fixed point $\beta(g_R^{\ast})=0$ when
\begin{equation}
g_R^{\ast}=\{2^{-d/2}B_2\Gamma(\epsilon/2)\}^{-1}. \label{fp}
\end{equation}
The fixed point is of order $\epsilon$. Finally, the expansion of $g_0$ in
powers $g_R$ remains, as in \cite{Lee}:
\begin{equation}
g_0=g_R + {g_R^2\over g_R^{\ast}}+\dots\quad \label{gsub}
\end{equation}

\subsubsection{Callan-Symanzik Equation}

We now write down the renormalisation group equation for $\langle\phi_1
\rangle_R$ (the renormalised value of $\langle\phi_1\rangle$), expressing its
lack of dependence on the normalisation scale:
\begin{equation}
\left( \kappa{\partial \over \partial\kappa} +\beta(g_R){\partial\over\partial
g_R}\right)\langle\phi_1\rangle_R(x,g_R,\kappa,\bar J)=0.
\end{equation}
In addition, dimensional analysis implies
\begin{equation}
\left( \kappa{\partial\over\partial\kappa}-x{\partial\over\partial x}+(d+1)
\bar J{\partial\over\partial\bar J}\right)\langle\phi_1\rangle_R(x,g_R,
\kappa,\bar J)=d\langle\phi_1\rangle_R(x,g_R,\kappa,\bar J).
\end{equation}
Eliminating the terms involving $\kappa$, we have
\begin{equation}
\left( x{\partial\over\partial x}-(d+1)\bar J{\partial\over\partial\bar J}+
\beta(g_R){\partial\over\partial g_R}+d\right)\langle\phi_1\rangle_R(x,g_R
,\kappa,\bar J)=0.
\end{equation}
This can be solved by the method of characteristics, with solution
\begin{equation}
\langle\phi_1\rangle_R(x,g_R,\kappa,\bar J)=(\kappa x)^{-d}\langle\phi_1
\rangle_R(\kappa^{-1},\tilde g_R(\kappa^{-1}),\kappa,\tilde{\bar J}(\kappa
^{-1})), \label{csz}
\end{equation}
and associated characteristics
\begin{equation}
x{\partial \tilde g_R\over \partial x}=\beta(\tilde g_R) \qquad \tilde
g_R(x)=g_R
\end{equation}
\begin{equation}
x{\partial\tilde{\bar J}\over\partial x}=-(d+1)\tilde{\bar J} \qquad
\tilde{\bar J}(x)=\bar J.
\end{equation}
These equations have the exact solutions:
\begin{equation}
\tilde{\bar J}(x')=\left({x\over x'}\right)^{d+1}\bar J \label{runJ}
\end{equation}
\begin{equation}
\tilde g_R(x')=g_R^{\ast}\left( 1+{g_R^{\ast}-g_R\over g_R\left({x\over
x'}\right)^{\epsilon}}\right)^{-1}, \label{rung}
\end{equation}
where in the large $|x|$ limit $\tilde g_R\rightarrow g_R^{\ast}$.

We can make use of the mechanics developed above by first calculating an
expansion in powers of $g_0$, which can be converted into an expansion
in powers of $g_R$ via (\ref{gsub}). Provided that the expansion is
non-singular in $\epsilon$, we can relate the $g_R$ expansion to an
$\epsilon$ expansion using (\ref{csz}), where for large $|x|$ we can
take $\tilde g_R\rightarrow g_R^{\ast}$.

\subsubsection{Tree Diagrams}

At this point we need to identify the leading terms in an expansion in powers
of $g_0$ - something which can be done in a very similar fashion to
\cite{Lee}, using Field Theory I. For the calculation of $\langle\phi\rangle$,
tree diagrams are of
order $g_0^{i}\bar J^{1+i}$, for integer $i$. Diagrams with $j$ loops will be
of order $g_0^{i}\bar J^{1+i-j}$. As the addition of loops makes the power of
$g_0$ higher relative to the power of $\bar J$, we see that
the number of loops will give an indicator of the order of the
diagram.

We are now in a position to develop two tree level quantities - namely the
classical density and the classical response function. Diagrammatically, we
represent the classical densities by wavy lines and the classical response
functions by thick lines. The tree level density
$\langle\phi\rangle$ is given by the sum of all tree diagrams which end
with a $G_{\phi\bar\phi}$ propagator, as shown in figure 4(b). This is
equivalent to the mean field equation, as may be seen by acting on both
sides of the graphical equation by the inverse Green function $2(\partial_{
\bar t}-\nabla^2)$.
Similarly, acting on the much simpler tree level diagram for
$\langle\psi\rangle$ (figure
4(a)) with
the inverse Green function gives its classical equation.

We now define the three response functions for the theory:
\begin{eqnarray}
& &\langle\psi(x,-k_{\perp},-s)\bar\psi(x',k_{\perp},s)\rangle^{(1)}\qquad
\langle
\phi(x,-k_{\perp},-s)\bar\phi(x',k_{\perp},s)\rangle^{(1)}
\nonumber\\ & &\qquad\qquad\qquad\langle\phi
(x,-k_{\perp},-s)\bar\psi(x',k_{\perp},s)\rangle^{(1)},
\end{eqnarray}
where the superscript `$1$' indicates that they are defined in Field Theory I.
Their diagrammatic sums are shown
in figure 5. The first one: $\langle\psi(x,-k_{\perp},-s)\bar
\psi(x',k_{\perp},s)\rangle^{(1)}$ is simply the propagator $G_{\psi_1\bar\psi}
^{(2)}$, where the superscript `$2$' indicates that it belongs to the second
field theory.
It is also easy to show that the second response function
$\langle\phi(x,-k_{\perp},-s)\bar\phi(x',k_{\perp},s)\rangle^{(1)}$ is
equivalent to
the propagator $G_{\phi_1\bar\phi}^{(2)}$. To do this,
we rearrange the unrescaled equation for $G_{\phi_1\bar\phi}^{(2)}$:
\begin{equation}
2(s+k_{\perp}^2-\partial_x^2)G_{\phi_1\bar\phi}^{(2)}(x,x',k_{\perp},s)
=-4\bar\lambda\phi_c(x) G_{\phi_1\bar\phi}^{(2)}(x,x',k_{\perp},s)
+\delta(x-x').
\end{equation}
Including
a delta function integration in the first term on the right hand side, and
acting on both sides with the inverse Green function, we obtain
\begin{equation}
G_{\phi_1\bar\phi}^{(2)}(x,x',k_{\perp},s)=G_{\phi\bar\phi}^{(1)}
-4\bar\lambda\int G_{\phi\bar\phi}^{(1)}(x,x_1,k_{\perp},s)
\phi_c(x_1)G_{\phi_1\bar\phi}^{(2)}(x_1,x',k_{\perp},s)dx_1,
\end{equation}
where $G_{\phi\bar\phi}^{(1)}$ is the $\phi$ propagator for the first
field theory.
Iteration now generates the appropriate tree level expansion for
$\langle\phi(x,-k_{\perp},-s)\bar\phi(x',k_{\perp},s)\rangle^{(1)}$, and we
have $G_{\phi_1\bar\phi}^{(2)} =\langle\phi\bar\phi\rangle^{(1)}$.
The remaining response
function $\langle\phi\bar\psi\rangle^{(1)}$ is, as would
be expected, equivalent in the second field theory to the two point vertex
sandwiched between a $\phi$ propagator and a $\psi$ propagator.

\section{Density and Reaction Front Calculations}

We first note that we cannot draw diagrams which terminate with a
$\psi_1$ propagator in Field Theory II. Consequently, we conclude that
$\langle\psi_1\rangle=0$, and hence that
$\langle\psi\rangle=\psi_c$. This also follows from averaging equation
(\ref{ee1}). We now turn to the asymptotic evaluation of $\langle\phi\rangle$.
Inserting the classical (tree level) solution
(\ref{phic}) into the Callan-Symanzik solution (\ref{csz}), and making
the leading order replacements
$\bar\lambda\rightarrow g_R \kappa^{\epsilon}$, and for large $|x|$,
$\tilde g_R \rightarrow g_R^{\ast}$, we obtain
\begin{equation}
\langle\phi\rangle={1\over 2}\bar J |x|+0.3787{\bar J^{7/12}\over {g_R^{\ast}
}^{5/12}}|x|^{(7-5d)/12}e^{-(2/3){g_R^{\ast}}^{1/2}\bar J^{1/2}|x|^{(d+1)/2}}
+ \dots \end{equation}
If we use the explicit value of $g_R^{\ast}$ from (\ref{fp}), then
\begin{equation}
\langle\phi\rangle={1\over 2}\bar J |x|+0.3787{\bar J^{7/12}\over(4\pi
\epsilon)^{5/12}}|x|^{(7-5d)/12}e^{-(2/3)(4\pi\epsilon)^{1/2}\bar J^{1/2}
|x|^{(d+1)/2}} + \ldots
\end{equation}
So the tree level expression consists of the expected linear term, which
must be present if the boundary conditions are to be satisfied, together
with a stretched exponential component.

\subsection{One Loop Contributions}

According to our earlier arguments we expect the next order contributions
to $\langle\phi\rangle$ (in Field Theory I) to contain one loop embedded
somewhere in the tree
diagram. The diagrams corresponding to this prescription are shown in
figure 6. However, we have shown that we may translate these diagrams into
Field Theory II by replacing response functions by propagators. This is
convenient as we have analytic expressions for the Green functions in
the second field theory (at least asymptotically), and so performing
calculations becomes easier. The equivalent diagrams for Field Theory II are
shown in figure 7. Notice that the density lines present in the diagrams for
Field Theory I have been absorbed into the vertices for Field Theory II,
where a factor of $\phi_c^2-\psi_c^2$ is present at the source vertex.

We begin by calculating the loop contained in the third diagram of figure 7
(whilst not
including its leftmost vertex). At this level of
approximation, we replace the source (the incoming classical density lines
at the rightmost vertex in Field Theory I) by a delta function at the origin,
with a weight equal to the area under the classical reaction front; in other
words:
\begin{equation}
\phi_c^2-\psi_c^2\rightarrow \left(\int(\phi_c^2-\psi_c^2)dx'\right)
\delta(x).
\end{equation}
This will be valid provided $\langle\phi\rangle$ decays much more slowly
than the classical reaction front - an assumption that will be shown to be
justified {\it a posteriori}. By integrating the classical equations, we also
have the relation
\begin{equation}
\bar\lambda\int(\phi_c^2-\psi_c^2)dx'=\bar\lambda^{1/3}\bar J^{4/3}\int h[(\bar
\lambda\bar J)^{1/3}x']dx'=\bar J,
\end{equation}
which is simply saying that, classically, the number of particles entering
the system is the same as the number being annihilated at the reaction front.
This relationship is also true non-classically, if we average over the
noise. After we have performed the rescaling $\tilde x'= (\bar\lambda\bar J)
^{1/3}x'$, this becomes
\begin{equation}
\int h(\tilde x')d\tilde x'=1.
\end{equation}
So the vertex factor at the source becomes:
\begin{equation}
\bar\lambda^{1/3}\bar J^{4/3}h(\tilde x)\rightarrow\bar\lambda^{1/3}\bar J
^{4/3}\left( \int h(\tilde x')d\tilde x'\right) \delta(\tilde x)=\bar\lambda
^{1/3}\bar J^{4/3}\delta(\tilde x).
\end{equation}
Hence the loop is given by the integral
\begin{eqnarray}
& &2\bar\lambda^{-1/3}\bar J^{2/3}(\bar\lambda\bar J)^{d/3}\int {e^{-
(\tilde k_{\perp}^2+i\tilde s)^{1/2}|\tilde x_1''|}\over 4(\tilde k_{\perp}
^2+i\tilde s)^{1/2}}{e^{-(\tilde k_{\perp}^2-i\tilde s)^{1/2}|\tilde x_2''|}
\over 4(\tilde k_{\perp}^2-i\tilde s)^{1/2}}
{e^{-(\tilde x'+\tilde k_{\perp}^2+i
\tilde s)^{1/2}|\tilde x'-\tilde x_1''|}\over 4(\tilde x'+\tilde k_{\perp}
^2+i\tilde s)^{1/2}}\nonumber \\ & & \qquad \qquad \qquad \qquad \qquad
{e^{-(\tilde x'+\tilde k_{\perp}^2- i\tilde s)^{1/2}|
\tilde x'-\tilde x_2''|}\over 4(\tilde x'+\tilde k_{\perp}^2-i\tilde s)^
{1/2}}(2\tilde x_1'')(2\tilde x_2'')d\tilde
x_1''d\tilde x_2''{d^{d-1}\tilde k_{\perp}d\tilde s\over (2\pi)^d},
\end{eqnarray}
where the prefactor of `$2$' counts the
number of possible diagram configurations, and the $s$ integration is along the
real axis. In the integral we have used the form of the propagator for the
$\phi$ field valid for $\tilde x_1'',\tilde x_2'',\tilde x' \gg0$, and
$\tilde x_1'',\tilde x_2''\approx \tilde x'$, the
region from which we
expect the dominant contribution (as here the $\phi$ propagator falls off only
as a power law). The $\tilde x_1''$ and $\tilde x_2''$ integrations are
elementary, giving
\begin{eqnarray}
& & 2\bar\lambda^{-1/3}\bar J^{2/3}(\bar\lambda\bar J)^{d/3}{1\over 16}\int
e^{-(\tilde k_{\perp}^2+i\tilde s)^{1/2}\tilde x'}e^{-(\tilde k_{\perp}^2-i
\tilde s)^{1/2}\tilde x'}\left[(\tilde k_{\perp}^2+i\tilde s)^{-1/2}-{2\over
\tilde x'^2}\right] \nonumber \\ & & \qquad \qquad \qquad \qquad \qquad\qquad
\left[(\tilde k_{\perp}^2
-i\tilde s)^{-1/2}-{2\over \tilde x'^2}\right]{d^{d-1}\tilde k_{\perp}d
\tilde s\over (2\pi)^d}.
\end{eqnarray}
However, we notice that the leading $\tilde x'$ part of the integral
is in fact a divergent power law. Furthermore, this divergence cannot
be cancelled by the renormalisation of the theory, as any such
cancellation would have to arise from coupling constant renormalisation
at the tree level (using (\ref{gsub})). As the renormalised tree
level result is still an
exponential, cancellation with a power law cannot occur. Consequently,
we must find another mechanism for the removal of the divergence, and
this is provided by its cancellation with
the divergent loop shown in figure 7b. Turning now to the next to
leading $\tilde x'$ term
in the above integral, we have
\begin{eqnarray}
& &-\bar\lambda^{-1/3}\bar J^{2/3}(\bar\lambda\bar J)^{d/3}{1\over
4\tilde x'^2}
\int e^{-(\tilde k_{\perp}^2+i\tilde s)^{1/2}\tilde x'}e^{-(\tilde k_{\perp}
^2-i\tilde s)^{1/2}\tilde x'}\nonumber\\ & &\qquad\qquad \left[{1\over
(\tilde k_{\perp}^2+i\tilde s)^
{1/2}}+{1\over (\tilde k_{\perp}^2-i\tilde s)^{1/2}}\right]{d^{d-1}\tilde k_
{\perp}d\tilde s\over (2\pi)^d}.
\end{eqnarray}
This can be rewritten as
\begin{equation}
\bar\lambda^{-1/3}\bar J^{2/3}(\bar\lambda\bar J)^{d/3}{2\over\tilde x'^2}{
\partial\over\partial \tilde x'}\int {2e^{-(\tilde k_{\perp}^2+i\tilde s)^{1/2}
\tilde x'-(\tilde k_{\perp}^2-i\tilde s)^{1/2} \tilde x'}\over 4(\tilde k_
{\perp}^2+i\tilde s)^{1/2}4(\tilde k_{\perp}^2-i\tilde s)^{1/2}}{d^{d-1}
\tilde k_{\perp}d\tilde s\over (2\pi)^d},
\end{equation}
i.e. a constant times the derivative of the $\psi$ propagator loop
integral. Rewriting the $\psi$ propagators entirely in momentum space, and
performing a contour integration for $\tilde s$, we end up with
\begin{equation}
\bar\lambda^{-1/3}\bar J^{2/3}(\bar\lambda\bar J)^{d/3}
{1\over \tilde x'^2}{\partial\over\partial\tilde x'}\int{e^{i\tilde p\tilde x'}
\over {\bf\tilde k}^2+(\tilde p-{\bf\tilde k})^2}{d^d{\bf\tilde
k}d\tilde p\over (2\pi)^{d+1}}.
\end{equation}
This integral may be done exactly using some standard results from \cite{GR},
with the result
\begin{equation}
\bar\lambda^{-1/3}\bar J^{2/3}(\bar\lambda\bar J)^{d/3}2^{-1-d}\pi^{-(d+1)/2}
(1-d)\Gamma\left({d-1\over 2}\right)\tilde x'^{-d-2}.
\end{equation}
To evaluate the contribution to $\langle\phi\rangle$ we now need to include the
leftmost vertex and propagator:
\begin{equation}
-(\bar\lambda\bar J)^{d/3}2^{-2-d}\pi^{-(d+1)/2}(1-d)\Gamma\left({d-1\over 2}
\right)\int\tilde x'^{-d-2}\tilde x^{-1/2}e^{-\tilde x^{1/2}|\tilde x-
\tilde x'|}d\tilde x',
\end{equation}
giving the leading order result
\begin{equation}
-(\bar\lambda\bar J)^{-1}2^{-1-d}\pi^{-(d+1)/2}(1-d)\Gamma\left({d-1\over 2}
\right)x^{-d-3}.
\end{equation}
We may now insert this into the Callan-Symanzik solution (\ref{csz}), and
use the results for the running current/coupling (\ref{runJ})/(\ref{rung}),
and for the coupling fixed point (\ref{fp}). This leads to the 1 loop density
correction
\begin{equation}
{x^{-2d-1}\over 32\pi^2\bar J\epsilon},
\end{equation}
which justifies the use of the delta function approximation for the source.
The one remaining diagram in figure 7 consists entirely of $\phi$
propagators, and so asymptotically we expect an exponential dependence
which we neglect in comparison with the power law. So to this level of
accuracy, we have
\begin{equation}
\langle\phi\rangle={1\over 2}\bar J|x|+0.3787{\bar J^{7/12}\over (4\pi
\epsilon)^{5/12}}|x|^{(7-5d)/12}e^{-{2\over 3}(4\pi\epsilon)^{1/2}\bar J
^{1/2}|x|^{(d+1)/2}}+{|x|^{-2d-1}\over 32\pi^2\bar J\epsilon} +\ldots
\label{1loop}
\end{equation}
Using the relation (\ref{ptr}) it is also straightforward to calculate the
form of the reaction front profile:
\begin{eqnarray}
& &R=\lambda\langle\phi^2-\psi^2\rangle=0.3787\bar J^{19\over 12}
{(4\pi\epsilon)^{7/12}\over
9/D}(d+1)^2|x|^{(7d-5)/12}e^{-{2\over 3}(4\pi\epsilon)^{1/2}\bar J
^{1/2}|x|^{(d+1)/2}} \nonumber \\ & &\qquad\qquad\qquad\qquad\qquad +{(2d+1)
(2d+2)|x|^{-2d-3}\over 32\pi^2\epsilon\bar J/D} + \ldots, \label{RScale1}
\end{eqnarray}
where only the leading terms generated by the differentiation of each
of the component parts of (\ref{1loop}) have been retained.

Finally, we consider the cancellation of divergences at 1 loop which,
as we mentioned earlier, can most easily be seen in the formalism
of Field Theory I. We expect divergent contributions from the 1 loop diagrams
a,b,c and d in figure 6, in the limit where the position of the loop's left
vertex tends towards that of the right vertex. In this limit, where no
insertions are possible into the response functions, it is
appropriate to replace the loop of $\phi$ response functions with one of
$\psi$ response functions. Evaluating this loop gives the result
$1/2g_R^{\ast}$, and the diagrams become as shown in figure 8.
However, if we consider the corrections to the tree level due to subleading
terms in $g_0(g_R)$ (from (\ref{gsub})), we have the same diagrams but with
opposite signs, which exactly cancel the 1 loop divergences.

\subsubsection{Two Loops}

Whilst we have not calculated in full the contributions to the
density from the two loop diagrams, a remark concerning their general
nature, and of the nature of our perturbation expansion, is in order. A
sample of these two loop diagrams is shown in figure 9.

The easiest diagrams to evaluate are the first and second of those in figure
9, for which it is easy to check that they have the form
\begin{equation}
\sim {|x|^{-4d-3}\bar J^{-3}\over\epsilon^2}\sim\epsilon^{-1/2}|x|^{-d}[
\bar J|x|^{d+1}\epsilon^{1/2}]^{-3}.
\end{equation}
Hence the perturbative expansion for the power law contributions to
$\langle\phi\rangle$ would appear to have the form:
\begin{equation}
\langle\phi\rangle\sim\epsilon^{-{1\over 2}}|x|^{-d}[\bar J|x|^{d+1}\epsilon
^{1/2}]+\epsilon^{-{1\over 2}}|x|^{-d}[\bar
J|x|^{d+1}\epsilon^{1/2}]^{-1}+\epsilon
^{-{1\over 2}}|x|^{-d}[\bar J|x|^{d+1}\epsilon^{1/2}]^{-3}+\dots
\end{equation}
Consequently, we see that the condition for our field theory to be valid is
that the dimensionless parameter $\bar J|x|^{d+1}$ be $\gg 1$.
It should also be noted that subleading power laws from the loop
integrals will be smaller than their leading term by factors of
$(\bar J|x|^{d+1})^{-1/3}$.

\subsection{$d\geq 2$}

At the upper critical dimension for the system, in this case $d=2$, we expect
logarithmic corrections to the $d>2$ results, owing to the presence of
the marginally irrelevant parameter $\bar\lambda$. The Callan-Symanzik
solution (\ref{csz}) is still valid, although with a different coupling, which
we calculate by taking $\epsilon\rightarrow0$ in (\ref{bf}). This gives the
running coupling
\begin{equation}
\tilde g_R(\kappa^{-1})={g_R\over 1+{g_R\over 4\pi}\ln(\kappa x)}.
\end{equation}
The behaviour of the running current is as previously calculated. Using the
asymptotic form $\tilde g_R\sim 4\pi/\ln(\kappa x)$, we obtain
\begin{equation}
\langle\phi\rangle={1\over 2}\bar J|x|+0.3787\bar J^{7\over 12}\left({\ln
|x|\over 4\pi}\right)
^{5\over 12}|x|^{-{1\over 4}}e^{-{2\over 3}(4\pi\bar J)^{1/2}(\ln |x|)
^{-1/2}|x|^{3/2}}
+{|x|^{-5}\ln |x|\over 32\pi^2\bar J} +\ldots ,
\end{equation}
where higher order corrections will be only $O[(\ln |x|)
^{-1}]$ smaller, so
the asymptotic regime will be accordingly hard to reach. Finally, for the
reaction front, we have
\begin{equation}
R=0.3787D{(4\pi)^{7/12}\bar J^{19/12}
\over
(\ln |x|)^{7/12}}|x|^{3/4}e^{-{2\over 3}(4\pi\bar J)^{1/2}(\ln |x|)^{-1/2}|x|
^{3/2}} +{15D\bar J^{-1}|x|^{-7}\ln |x|\over 16\pi^2} + \ldots
\end{equation}

For dimensions higher than the critical dimension, the expressions from the
evaluation of the Feynman diagrams are used directly without being inserted
into the Callan-Symanzik solution. This gives us the results, valid
for $d>2$, and in the regime $(J^2\lambda/D^3)|x|^{d+4}\gg 1$:
\begin{eqnarray}
& &\langle\phi\rangle={1\over 2}\bar J|x|+0.3787{\bar J^{7/12}\over\bar\lambda
^{5/12}}|x|^{-1/4}e^{-{2\over 3}(\bar\lambda\bar J)^{1/2}|x|^{3/2}}\nonumber\\
& & \qquad\qquad-(\bar\lambda\bar J)^{-1}2^{-1-d}\pi
^{-(d+1)/2}(1-d)\Gamma\left({d-1\over 2}\right) |x|^{-d-3} + \ldots
\end{eqnarray}
and
\begin{eqnarray}
& &R=0.3787\bar J^{19/12}\bar\lambda^{7/12}
D|x|
^{3/4}e^{-{2\over 3}(\bar\lambda\bar J)^{1/2}|x|^{3/2}} \label{RScale2} \\ & &
-D\bar\lambda^{-1}\bar J^{-1}2^{-1-d}
\pi^{-(d+1)/2}\Gamma\left({d-1\over 2}\right)(1-d)
(d+3)(d+4)|x|^{-d-5} + \ldots \nonumber
\end{eqnarray}

\section{Interface Fluctuations}

We now turn to the related problem of the nature of fluctuations
in position of the reaction front. This is similar to the
question of the fluctuations of an interface in the dynamical Ising Model, as
described by the time dependent Landau-Ginzburg (TDLG) equation with
noise (for example in
model A - see \cite{HH}). This equation may be mapped to a path
integral for the field $\Phi$, with the introduction of response fields
$\tilde\Phi$, using the Martin-Siggia-Rose formalism:
\begin{equation}
\int{\cal D}\Phi{\cal D}\tilde\Phi e^{-\int dx d^{d-1}y dt[\tilde\Phi\{\dot
\Phi+\Gamma(\nabla^2\Phi+V'(\Phi))\}+{1\over 2}\Gamma\tilde\Phi^2]},
\end{equation}
where the last term in the action results from averaging over the noise.
Solving the TDLG
equation in the absence of noise gives us the
classical profile $\Phi_c$, and on physical
grounds we expect the full functional form of $\Phi$ to be $\Phi_c(x-f(y,t))
\approx \Phi_c(x)-f(y,t)\Phi'_c(x)$. The idea now is to substitute
this into the action and to
expand the response fields in terms of some complete set of
eigenfunctions $\Psi_n(x)$ :
\begin{equation}
\tilde\Phi(x,y,t)=\sum_n A_n\tilde f_n(y,t)\Psi_n(x),
\end{equation}
where the $\{A_n\}$ are normalising constants. This set is chosen
such that when the $x$ dependence is integrated out of the action, it leaves
behind an unambiguous equation for $f(y,t)$, obtained by integrating
over the new response fields $\tilde f(y,t)$ in the path integral. For
the Ising case, $f(y,t)$ can be shown to satisfy a noisy diffusion
equation, whose solution implies that fluctuations delocalise the
interface for $d\leq 3$. A similar result for reaction fronts would
have dramatic consequences.

Returning to the reaction-diffusion system, we
expect the functional forms for the fields in our geometry
to be $\psi(x-f_1(y,t))$ and $\phi(x-f_2(y,t))$, by analogy with the
Ising case.
Considering first the situation where we neglect noise in the system, we
expand the above functional forms, giving
\begin{eqnarray}
& &\phi\approx\phi(x)-f_2(y,t)\phi'(x) \Rightarrow \phi={1\over 2}
\bar J|x|\mp{1
\over 2}\bar J f_2 \qquad (x=\pm L) \\
& &\psi=\psi(x)-f_1(y,t)\psi'(x) \Rightarrow \psi=-{1\over 2}\bar Jx+{1
\over 2}\bar J f_1.
\end{eqnarray}
Hence $a=(1/2)\bar J(f_1-f_2)$ and $b=\bar Jx-(1/2)\bar J(f_1+f_2)$ at $x=L$,
and $a=-\bar Jx+(1/2)\bar J(f_1+f_2)$ and $b=(1/2)\bar J(f_2-f_1)$ at $x=-L$.
In the absence of noise $a$ and $b$ represent the (positive) particle
densities, so we must have $f_1=f_2$.

However, if we include the noise term then this argument is invalid, and we
proceed, as in the Ising case, by inserting the expanded functional
forms for $\phi$ and $\psi$ into the action for Field Theory I, giving
\begin{eqnarray}
& & S=\int dx d^{d-1}y dt[2\bar\phi\{\phi_c'[-\dot
f_2+\nabla_{\perp}^2 f_2]+2\bar
\lambda\psi_c\psi_c'(f_1-f_2)\} \nonumber\\ & &\qquad\qquad\qquad\qquad
+2\bar\psi\{\psi_c'[-\dot
f_1+\nabla_{\perp}
^2 f_1]\}+\bar\lambda(\bar\phi^2-\bar\psi^2)(\phi_c^2-\psi_c^2)],
\end{eqnarray}
where we have made the approximation $\phi\rightarrow\phi_c$, and then
used the classical equations to simplify the expression. Here $y$ are
the coordinates for directions perpendicular to the applied currents.
In our case it is now appropriate to Fourier expand the $\bar\psi$ and
$\bar\phi$ fields, i.e.
\begin{equation}
\bar\psi=\sum_n\zeta_n(y,t)\theta_n(x) \qquad \qquad \bar\phi=\sum_n \xi_n
(y,t)\theta_n(x),
\end{equation}
where $\theta_n=\sin(n\pi x/L)$ for $n>0$, $\theta_n=1/2$ for $n=0$, and
$\theta_n=\cos(n\pi x/L)$ for $n<0$. Inserting this into the noise term and
performing the $x$ integration, we have
\begin{equation}
\int\bar\psi^2(\phi_c^2-\psi_c^2)dx=\sum_{n,m}\zeta_n\zeta_m\int_{-L}^
{L}\theta_n\theta_m(\phi_c^2-\psi_c^2)dx=\sum_{n,m}\zeta_n X_{nm}\zeta_m,
\label{a}
\end{equation}
where $X_{nm}$ is a symmetric matrix which we now diagonalise. Using
$\hat\zeta_n=D_{nm}\zeta_m$, but such that $\hat\zeta_0=\zeta_0$, we
rewrite (\ref{a}) as
\begin{equation}
\sum_{n,m}\hat\zeta_n\Lambda_{nm}\hat\zeta_m,
\end{equation}
where ${\bf\Lambda}$ is a diagonal, and ${\bf D}$ a diagonalising, matrix.
Bearing in mind the symmetries of the classical solutions, we can perform
the $x$ integration within the action to arrive at the path integral:
\begin{eqnarray}
& & \int{\cal D}f_1{\cal D}f_2\prod_n{\cal D}\hat\zeta_n{\cal D}\xi_n
\exp\left[-\int d
^{d-1}y dt\left\{2\sum_{n>0}\xi_n[A_n(-\dot f_2+\nabla_{\perp}^2f_2)
\right.\right.\\ & &\left.\left. +2\bar\lambda
B_n(f_1-f_2)]+\bar\lambda\sum_{n,m}
\xi_n\xi_mX_{nm}+2\hat\zeta_0 C_0[-\dot f_1
+\nabla_{\perp}^2f_1]-\bar\lambda\sum_n\hat\zeta_n\Lambda_{nn}\hat\zeta_n
\right\}\right], \nonumber
\end{eqnarray}
with
\begin{equation}
A_n=\int_{-L}^L\theta_n\phi_c'dx,\qquad B_n=\int_{-L}^L\theta_n\psi_c\psi_c'dx
, \qquad C_0=\int_{-L}^L\theta_0\psi_c'dx\sim \bar JL.
\end{equation}
Integrating over $\hat\zeta_0$ and $f_1$ gives the equation
\begin{equation}
-\dot f_1+\nabla_{\perp}^2 f_1+\eta=0, \label{b}
\end{equation}
where $\eta$ is a (possibly imaginary) noise variable, with a Gaussian
distribution:
\begin{equation}
P(\eta)\sim e^{-(\eta^2 C_0^2/\bar\lambda|\Lambda_{00}|)}.
\end{equation}
If we also diagonalise the noise term involving the $\xi$ fields, then the
relevant part of the action is transformed to
\begin{equation}
2\sum_{n>0}W_n\hat\xi_n[A_n(\dot g-\nabla_{\perp}^2 g-\eta)-2\bar\lambda B_n
g]+\bar\lambda\sum_n\hat\xi_n\Lambda_{nn}\hat\xi_n,
\end{equation}
where $g=f_1-f_2$, and the equation for $f_1$ (\ref{b}) has been added into
the action. The $\{W_n\}$ are coefficients generated by writing $\xi_n$ in
terms of $\{\hat\xi_m\}$. Finally, performing the integrations over $\hat
\xi_n$ and $g$, we find equations for $g$ which can only be mutually
consistent for different $n$ if $g=0$, or in other words, if $f_1=f_2=f$.
{}From (\ref{a}) we see that $\Lambda_{00} \sim \bar J\bar\lambda^{-1}$, and so
\begin{equation}
-\dot f+\nabla_{\perp}^2 f+\eta=0,
\end{equation}
where $\eta$ is a Gaussian noise variable with probability
distribution:
\begin{equation}
P(\eta)\sim e^{-(const.)L^2 \bar J\eta^2}.
\end{equation}
We now proceed to calculate the mean square fluctuation $\langle f^2
\rangle-\langle f\rangle ^2$. This can be done in a straightforward manner,
solving the noisy diffusion equation satisfied by $f$ using a Green function
method. The results are
\begin{equation}
\langle f(y,t)^2\rangle-\langle f(y,t)\rangle ^2 \sim \cases{{\Lambda^{d-3}
\bar J^{-1}\over L_{\parallel}^2} & for $d\geq 4$\cr {\ln(L_{\perp}\Lambda)
\bar J^{-1}\over L_{\parallel}^2} & for $d=3$\cr {L_{\perp}\bar J^{-1}\over
L_{\parallel}^2}& for $d=2$,\cr}
\end{equation}
where the system has physical dimensions $L_{\parallel}\times L_{\perp}^{d-1}$,
and $\Lambda$ is now the large $k$ momentum cut-off.
So we expect that interface fluctuations will be unimportant if
\begin{equation}
\langle f^2\rangle-\langle f \rangle ^2\sim
\cases {{\Lambda^{d-3}\over L_{\parallel}^2 JD^{-1}}\ll L_{\parallel}^2 \qquad
\Rightarrow \qquad {L_{\parallel}^2(JD^{-1})^{1/2}\over\Lambda^{(d-3)/2}}\gg
1 & for $d\geq 4$\cr
{\ln (L_{\perp}\Lambda)\over L_{\parallel}^2 JD^{-1}}\ll L_{\parallel}^2
\qquad\Rightarrow\qquad {L_{\parallel}^2(JD^{-1})^{1/2}\over(\ln (L_{\perp}
\Lambda))^{1/2}}\gg 1 & for $d=3$ \cr
{L_{\perp}\over L_{\parallel}^2 JD^{-1}}\ll L_{\parallel}^2 \qquad\Rightarrow
\qquad{L_{\parallel}^2(JD^{-1})^{1/2}\over L_{\perp}^{1/2}}\gg 1 & for $d=2$.}
\label{flca}
\end{equation}

These results can now be applied to the problem of the late time
behaviour of an initially homogeneous distribution of $A$ and $B$ particles
[23--26], where it has been shown that the reactants
segregate asymptotically \cite {BL,Lee3}. We assume here that we
can access the quasistatic time dependent regime by simply replacing
our currents $J$ by their time dependent analogues (this point is
discussed further in the next section). In \cite{Lee3} it is
demonstrated that these time dependent inward currents (towards
the domain interfaces) scale as $J\sim t
^{-(d+2)/4}$, where the domains have a characteristic length scale which
grows in time as $t^{1/2}$. So on the basis of our assumption
we can insert the appropriate time dependencies into (\ref{flca}),
{}from where it is easily seen that
fluctuations are unimportant for large enough $t$, in dimensions where
segregation occurs ($d<4$).

\section{Discussion}

The main results of our earlier calculations are expansions for the asymptotic
behaviour of the
density and reaction front profiles for dimensions above, below, and equal to
the critical dimension. We now compare our analytic results with
the available data from recent numerical simulations
\cite{DC,LAHS,CD,ALHS,C1}. Note that in all of these papers except \cite{CD},
the initial conditions are those of complete particle
segregation - so the particle currents at later times are time
dependent. The remaining reference \cite{CD} contains the results of
simulations in the steady state. In principle,
the calculations of this paper can be redone for the time dependent
case, but simple one loop considerations for $\langle\psi^2\rangle$
indicate that the dominant contributions to the integrals originate
{}from large times. At these times the reaction front is formed
quasistatically, and so we expect to be able to relate to the steady
state case by making the correspondence $J\sim t^{-1/2}$
\cite{C1,Lee3} (but see below for occasions where this breaks down).
Data for $d=2$ in the time dependent situation is presented in
\cite{DC} and \cite{LAHS}, although in \cite{DC} there is insufficient
information to extract the asymptotic behaviour of the reaction front.
Further simulations for $d=2$ and also for $d=1,3$ are given in
\cite{CD}, where evidence for (\ref{scale}) - their proposed scaling
form of $R$ - is given. The reaction front profile is seen to exhibit good
scaling collapse close to its centre for $d=1,2,3$ but again no
information is available for the asymptotics addressed in this paper.

Turning now to the $1d$ case, the simulations in \cite{ALHS,C1}
were performed using an infinite reaction rate constant, i.e.
if two particles of different species either crossed or occupied the
same lattice site, they immediately annihilated. With
initial conditions of complete particle segregation, this resulted in total
separation of the two species at all later times.
Consequently, the reaction front profile was determined by the
fluctuations in position of a delta function like reaction front.
Our results are for finite reaction rates, and are dominated by density
fluctuations which propagate out from the
reaction front centre to positions far away, a process which
cannot occur in the $1d$ simulations mentioned above. We
believe this to be the reason
for the discrepancy between our analytic calculations and the
numerical results. For
example, the $1d$ simulations of Cornell
produce evidence for a Gaussian reaction front profile, most notably
in figure 8 of \cite{C1}. In that graph $\log R$ is plotted against
$(x/X^{(2)})^2$, where $X^{(2)}$ is the width of the reaction {\it product}
profile $C=\int R(x,t) dt$, as measured by its second spatial moment.
The resulting straight line indicates that
the Gaussian profile is maintained well into the asymptotic region
(i.e. at least as far as $(x/X^{(2)})^2\approx 30$) - in exactly the
region where, in our model, we would expect our asymptotic expansion
to begin to apply.

In addition, controversy still exists over the spatial moments of
the reaction front profile - Araujo {\it et al} \cite{ALHS}
and Cornell \cite{C1} disagree over the presence of
multiscaling. In fact, our calculations suggest that multiscaling does
indeed occur for high enough moments in the time dependent version of
our model, starting from completely segregated initial conditions. For
the steady state situation,
the existence of the asymptotic power laws found above implies that
the moments:
\begin{equation}
x^{(q)}=\left({\int_{-\infty}^{\infty} |x|^{q} R(x) dx\over
\int_{-\infty}^{\infty} R(x) dx}\right)^{1/q} \label{mom}
\end{equation}
do not exist for $q\geq\omega+1$. However, in the time
dependent case, these moments {\it must} exist due to the presence of a
diffusive cutoff at $x_D\sim t^{1/2}$ \cite{CKD}. Therefore, for the
calculation of the spatial moments $x^{(q)}$ (for large enough $q$), we
cannot relate the steady state case to the time dependent case by
simply applying the scaling substitution $J\sim t^{-1/2}$.
We can make these remarks more quantitative by performing the
calculation of the spatial moments in the time dependent
situation. Separate arguments must be applied for $d<2$, when our RG
arguments imply that the steady state profile has a scaling form
(\ref{RScale1}); and
for $d>2$, when (\ref{RScale2}) shows that the fluctuation induced
power law tails do not scale. For $d<2$, we have:
\begin{equation}
x^{(q)}(t)\sim\left({\int_{-\infty}^{\infty} |x|^q t^{-\beta}S\left({x\over
t^{\alpha}}\right) F\left({x\over
t^{1/2}}\right) dx\over \int_{-\infty}^{\infty} t^{-\beta}S\left({x\over
t^{\alpha}}\right) F\left({x\over t^{1/2}}\right)
dx}\right)^{1/q}, \label{momtim}
\end{equation}
where $\alpha$ and $\beta$ are defined in the usual way \cite{GaRa,Lee3}.
Here $F(y)$ is a function which provides a
cutoff at $y\sim O(1)$, but whose inclusion does not affect
the calculation of moments which are finite even in the absence of a cutoff.
For $q<\omega+1$, where the $q$th moment of $S$ is finite (even
without a cutoff), we can therefore neglect the effects of $F$. However, for
$q>\omega+1$, the $q$th moment is infinite without the cutoff,
so the integral will now be dominated by the region $(x/t^{1/2})\sim
O(1)$, where the asymptotic result $S\sim (x/t^{\alpha})^{-\omega-2}$
may be used.
These considerations lead to the result $x^{(q)}(t)\sim t^{\alpha_q}$,
where (neglecting any logarithmic corrections for $q=\omega+1$):
\begin{equation}
\alpha_q=\cases{ \alpha & for $q<\omega+1$ \cr {1\over 2}+
{(\omega+1)(\alpha-1/2)\over q} & for $q>\omega+1$.}
\end{equation}
Hence we have a cusp at $q=\omega+1$, above which $\alpha_q$ tends
towards $1/2$ for large $q$. Note that this value of $1/2$ is
specific to a diffusive cutoff of the form $F(x/t^{1/2})$. For $d=2$
we also expect logarithmic corrections to the
above power laws.

For $d>2$, we must carry out a slightly different calculation, as
although the classical (tree level) reaction front obeys scaling,
(\ref{RScale2}) reveals that the one loop power law correction does not.
However, for moments which exist without a cutoff, it turns out that
the classical terms still give the dominant contribution in the scaling
limit. For these terms we have, in the steady state case:
\begin{equation}
\int_{-\infty}^{\infty} |x|^q R_c(x,\lambda,J) dx\sim \lambda^{1/3}J^{4/3}
\int_{-\infty}^{\infty} |x|^q S_c[(\lambda J) ^{1/3}x] dx \sim
\lambda^{-q/3} J^{1-q/3}.
\end{equation}
However, for the non-scaling power law we must consider:
\begin{equation}
(\lambda J)^{-1}\int_{(\lambda J^2)^{-1/(d+4)}}^{\infty} {x^q dx\over
x^{d+5}}\sim \lambda^{-q/(d+4)}J^{1-2q/(d+4)},
\end{equation}
where we have imposed a lower cutoff in the integral, derived
{}from the expansion parameter
of the $d>2$ asymptotic series (\ref{RScale2}). Comparing the $J$
dependence of the two results above, we
see that the first of these will
dominate in the scaling limit $J\rightarrow 0$.
Substituting $J\sim t^{-1/2}$ and normalising,
we end up with $\alpha_q=\alpha=1/6$, for $q<d+4$. For the higher
moments ($q>d+4$) we need to introduce the cutoff function $F$, so
the integral will be dominated by the region $(x/t^{1/2})\sim O(1)$,
where we can use the asymptotic power law from (\ref{RScale2}):
\begin{equation}
x^{(q)}\sim {(\lambda t^{-1/2})^{-1}\over t^{-1/2}}\int_{0}^{\infty}
dx{x^q\over x^{d+5}}
F(x/t^{1/2}) \sim \lambda^{-1} t^{(q-2-d)/2}.
\end{equation}
Consequently, we have the result $x^{(q)}\sim
t^{\alpha_q}$, where (neglecting logarithmic corrections for $q=d+4$):
\begin{equation}
\alpha_q= \cases{ {1\over 6} & for $q<d+4$ \cr
{1\over 2}-{d+2\over 2q} & for $q>d+4$.}
\end{equation}
In this case we have a discontinuity at $q=d+4$,
a result of the
power law term being unimportant for $q<d+4$, but dominant for
$q>d+4$. Once again we stress that the limiting behaviour
$\alpha_q\rightarrow 1/2$ as $q\rightarrow\infty$ is dependent on the
diffusive form of the cutoff.

Thus, we predict the existence of multiscaling in the time dependent
case in qualitative agreement with Araujo {\it et al}, even though we
are considering a different model. In general, power law tails in the
steady state reaction front profiles should {\it always} lead to
dynamic multiscaling, whatever their origin. These arguments
are similar to those of Cornell {\it et
al.} \cite{CKD}, who find evidence for multiscaling in the
reaction $nA+mB\rightarrow\emptyset$ with $(n,m)\neq(1,1)$. However,
in that case the
solutions of the mean field rate equations already give power laws,
even without the addition of fluctuation effects.

Finally, we conclude that the available simulations are not directly
applicable
to our calculations of asymptotic power laws and multiscaling. However, if the
asymptotics could be reached in a model with a finite
reaction rate, our results should be amenable to numerical
tests. These might be easiest in $1d$ where the power law tail
should be most pronounced.

\qquad

\noindent{\bf Acknowledgments.}

\noindent The authors would like to thank B. Lee for a reading of the
manuscript. We also acknowledge financial support from the EPSRC.

\newpage
\listoffigures
\newpage
\begin{figure}
\begin{center}
\leavevmode
\vbox{
\epsfxsize=4in
\epsffile{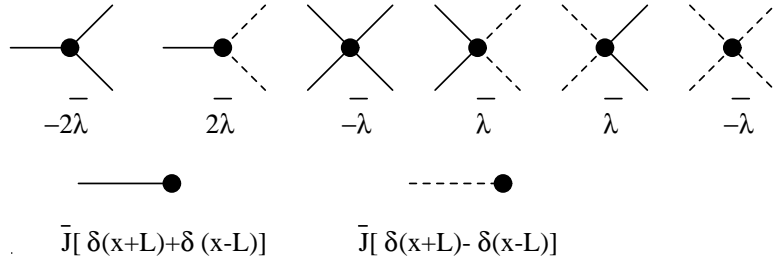}}
\end{center}
\caption{Vertices for Field Theory I.}
\end{figure}
\newpage
\begin{figure}
\begin{center}
\leavevmode
\hbox{%
\epsfxsize=5in
\epsffile{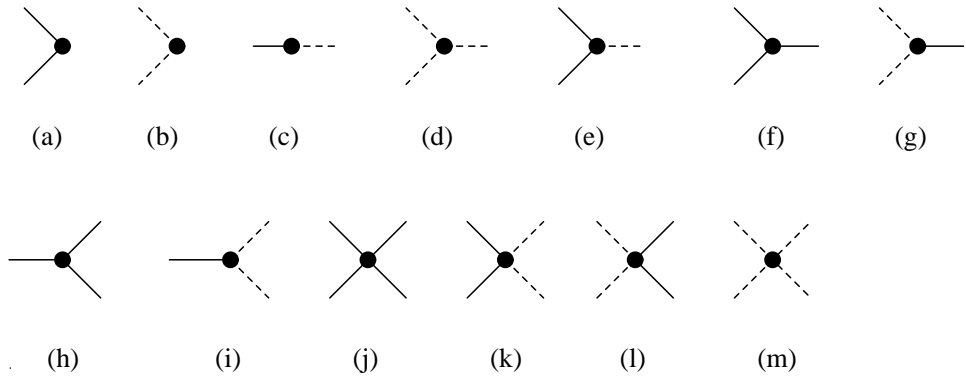}}
\end{center}
\caption{Vertices for Field Theory II. The
couplings associated with each of
the above diagrams are: (a). $-\bar\lambda^{1/3}\bar
J^{4/3}h(\tilde x)$ \quad (b).
$\bar\lambda^{1/3}\bar J^{4/3}h(\tilde x)$ \quad(c).
$-2(\bar\lambda\bar J)^{2/3}
\tilde x$ \quad (d). $(\bar\lambda\bar J)^{2/3}\tilde x$ \quad (e). $-(\bar
\lambda\bar J)^
{2/3}\tilde x$ \quad(f). $-2(\bar\lambda\bar J)^{2/3}f(\tilde x)$
\quad(g). $2(\bar
\lambda\bar J)^{2/3}f(\tilde x)$ \quad (h). $-2\bar\lambda$ \quad(i). $2\bar
\lambda$ \quad(j).
$-\bar\lambda$ \quad(k). $\bar\lambda$ \quad(l). $\bar\lambda$ \quad(m).
$-\bar\lambda$.}
\end{figure}
\newpage
\begin{figure}
\begin{center}
\leavevmode
\hbox{%
\epsfxsize=5in
\epsffile{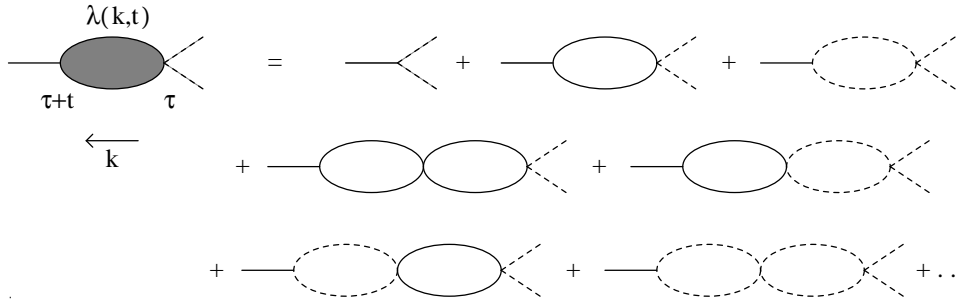}}
\end{center}
\caption{The sum of diagrams contributing to the primitively divergent
vertex function $\lambda(k,t)$.}
\end{figure}
\newpage
\begin{figure}
\begin{center}
\leavevmode
\hbox{%
\epsfysize=1.5in
\epsffile{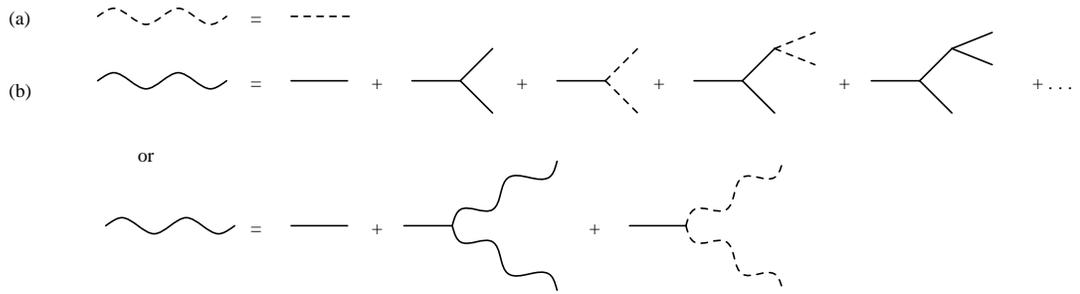}}
\end{center}
\caption{Tree level diagrams for (a) $\langle\psi\rangle$, and (b) $\langle
\phi\rangle$.}
\end{figure}
\newpage
\begin{figure}
\begin{center}
\leavevmode
\hbox{%
\epsfysize=1.55in
\epsffile{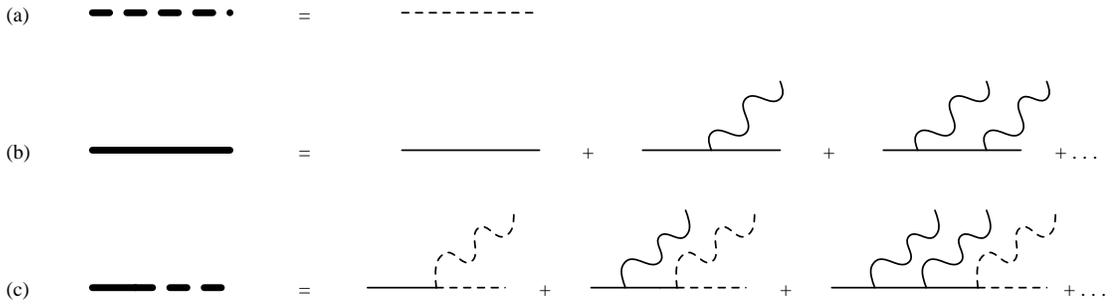}}
\end{center}
\caption{Response functions for Field Theory I:
(a) $\langle\psi\bar\psi
\rangle^{(1)}$, (b) $\langle\phi\bar\phi\rangle^{(1)}$, and (c)
$\langle\phi\bar\psi
\rangle^{(1)}$.}
\end{figure}
\newpage
\begin{figure}
\begin{center}
\leavevmode
\hbox{%
\epsfysize=2in
\epsffile{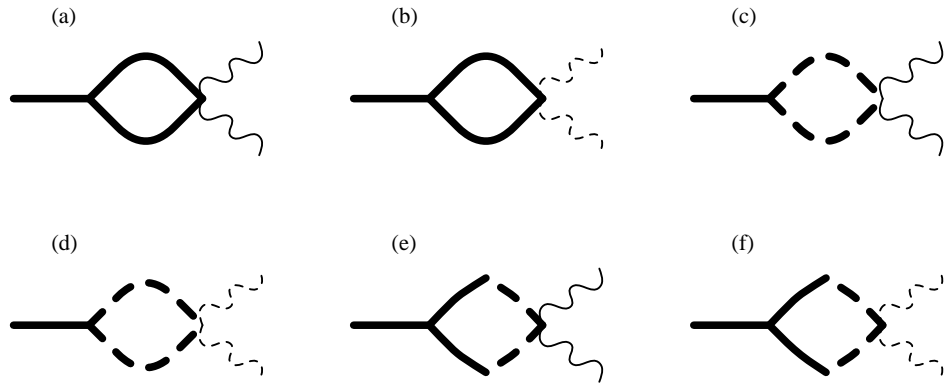}}
\end{center}
\caption{One loop diagrams in Field Theory I.}
\end{figure}
\newpage
\begin{figure}
\begin{center}
\leavevmode
\hbox{%
\epsfxsize=4.5in
\epsffile{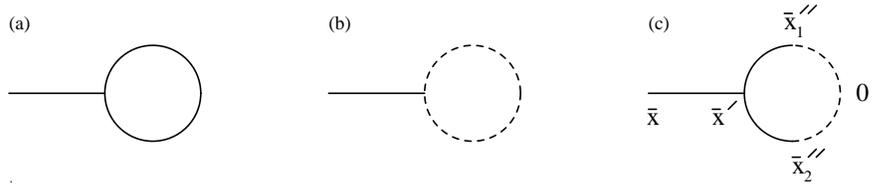}}
\end{center}
\caption{One loop diagrams in Field Theory II.}
\end{figure}
\newpage
\begin{figure}
\begin{center}
\leavevmode
\hbox{%
\epsfxsize=3in
\epsffile{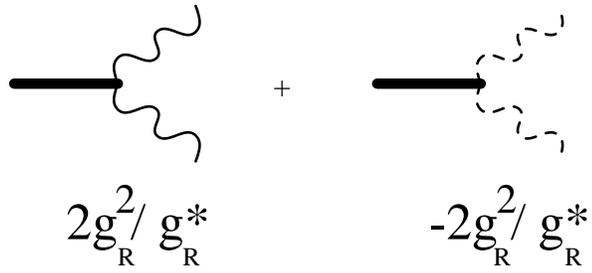}}
\end{center}
\caption{Divergences at one loop. The factor
underneath each diagram is associated with the vertex.}
\end{figure}
\newpage
\begin{figure}
\begin{center}
\leavevmode
\hbox{%
\epsfxsize=4.5in
\epsffile{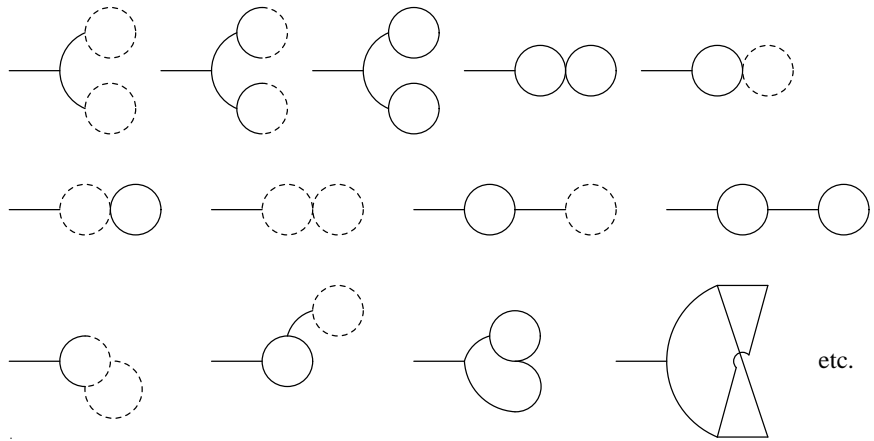}}
\end{center}
\caption{A sample of the two loop diagrams for
$\langle\phi_1\rangle$.}
\end{figure}
\end{document}